\documentclass[conference]{IEEEtran}
\IEEEoverridecommandlockouts

\usepackage{graphicx}
\usepackage{subcaption}
\usepackage{amsmath}
\usepackage{amssymb}
\usepackage{textcomp}
\usepackage{color}

\DeclareMathOperator*{\argmax}{arg\ max}
\DeclareMathOperator*{\diag}{diag}

\usepackage[makeroom]{cancel}

\usepackage{algorithm}
\usepackage{algpseudocode}

\usepackage[noadjust]{cite}

\usepackage{mathrsfs}

\usepackage{soul}

\usepackage{bm}

\newcommand*{\prob}{\mathsf{P}}

\begin{document}

\title{Chance-constrained optimal location of damping control actuators under wind power variability
\thanks{This material is based on work supported by the National Science Foundation under Grant No. 1509114. This work is also supported by the Engineering Research Center Program of the National Science Foundation and the Department of Energy under NSF Award No. EEC-1041877 and the CURENT Industry Partnership Program.}
}

\author{\IEEEauthorblockN{Horacio Silva-Saravia$^1$, H\'{e}ctor Pulgar-Painemal$^1$, Russell Zaretzki$^2$}
\IEEEauthorblockA{Department of Electrical Engineering and Computer Science$^1$, Business Analytics \& Statistics$^2$\\
University of Tennessee, Knoxville, TN, 37996\\
Email: hsilvasa@vols.utk.edu, hpulgar@utk.edu, rzaretzk@utk.edu }}

\maketitle

\begin{abstract}
This paper proposes a new probabilistic energy-based method to determine the optimal installation location of electronically-interfaced resources (EIRs) considering dynamic reinforcement under wind variability in systems with high penetration of wind power. The oscillation energy and total action are used to compare the dynamic performance for different EIR locations. A linear approximation of the total action critically reduces the computational time from hours to minutes. Simulating an IEEE-39 bus system with 30\% of power generation sourced from wind, a chance-constrained optimization is carried out to decide the location of an energy storage system (ESS) adding damping to the system oscillations. The results show that the proposed method, selecting the bus location that guarantees the best dynamic performance with highest probability, is superior to both traditional dominant mode analysis and arbitrary benchmarks for damping ratios.

\end{abstract}

\begin{IEEEkeywords}
oscillation energy, inter-area oscillations, small-signal stability, chance-constrained optimization, energy storage, renewable energy, wind power, random variable.
\end{IEEEkeywords}

\IEEEpeerreviewmaketitle

\section{Introduction}

The increasing penetration of variable renewable energy (VRE) into power systems, such as  solar and wind, creates operational and technical challenges. Among the technical challenges, low frequency oscillations are particularly affected. The causes include (1) the reduction of relative inertia added by electronically-interfaced generation and (2) the variability of the injected power, which creates power imbalance disturbances and unusual power flows. These power flows determine new uncommon equilibrium points, affecting power system dynamics and deteriorating small signal stability.

Traditionally, small signal stability analysis with sources of uncertainty has been modeled with probabilistic approaches by using eigenvalue sensitivities. Linear approximation of the relationship between system eigenvalues and a random system element, such as load or generation, allows one to characterize the distribution of system eigenvalues, given the probability density function (pdf) or, more generally, the distribution function of the random variables of interest \cite{Burchett1978}. Analytical methods have been investigated to construct the pdf of critical modes, determining the probability of system instability under scenarios of wind generation \cite{Bu2012}. Recently, nonlinear relationships have also been analyzed to obtain the cumulative distribution function (cdf) of the damping ratio of a dominant mode \cite{Wang2016}. The stochastic information of system eigenvalues can be used to control the system, for example, through optimal tunning of actuators by defining probabilistic objective functions of all system eigenvalues \cite{Bian2016}. Although these analyses provide tools to study and control specific operational requirements of a system, acceptable damping ratios for instance, these requirements, together with the study of a dominant mode, are usually arbitrary and fall short by not guaranteeing optimal system performance. From a practical perspective, we find a lack of a probabilistic analysis that considers a system's overall dynamic performance and creates an index that combines all system modes in a meaningful way.

The use of energy functions is attractive to provide a meaningful study of system oscillations. By analyzing the kinetic energy oscillation, a weighted combination of the system eigenvalues can be obtained. This weighted combination of eigenvalues has been used to determine energy exchange paths \cite{Jing1996,Yu2016} and can be defined as a system performance index for control \cite{Silva2017}. In particular, this kind of energy-based index can be utilized to solve a new problem related to the planning of the deployment of electronically-interfaced resources (EIR) such as non-conventional renewable generating systems (NRGS) or energy storage systems (ESS). The problem consists on finding the optimal location in the system to install EIR or ESS for dynamic reinforcement planning (DRP), i.e., damping electro-mechanical oscillations \cite{Pulgar2013,Silva2016,Poolla2017}. The optimal location has been studied by deterministic analysis and by relating system inertia distribution and residue index \cite{Pulgar2017,Yajun2017}. However, the study assumes a single very low damped inter-area mode, which is uncommon in large scale systems, like in the U.S. where the existence of several critical modes with similar damping ratios is usual. Unlike this approach, an energy-based system performance index has been shown to be suitable for capturing the overall system dynamic behavior considering all system eigenvalues \cite{Silva2017}. This index is based on the concept of total action and the optimal location is found by minimizing the area under the oscillation energy curve for a given disturbance; this dependency on the disturbance can be overcome by using a probabilistic framework.

This paper provides a new probabilistic measure to determine the optimal location of EIR in a DRP problem considering variability in systems with high penetration of wind power. The problem is solved by chance-constrained optimization, using a linear estimation of the total action as a system performance index in the objective function, and the probability of a set of disturbances. Simulations are performed in the IEEE-39 bus system with 30\% of power derived from wind penetration. Monte Carlo simulations show the computational advantage of the linear approximation compared to the exact calculation for analyzing the location of an ESS. The optimization results show that the best location for ESS differs from that obtained with traditional stochastic analysis of the dominant mode or an arbitrary benchmark for damping ratios. This occurs because the proposed approach chooses the location that maximizes the probability that the best system dynamic behavior is guaranteed. The paper is structured as follows. Section II describes the oscillation energy, the total action and its linear estimation. Section III formulates the chance-constrained optimization problem. Simulation results are presented in Section IV. Finally, Section V provides the conclusions.

\section{Oscillation energy analysis}

\subsection{Oscillation energy and action}

Electromechanical oscillations occur as a result of the kinetic energy exchange between different group of synchronous generators. The kinetic energy of each machine will oscillate depending the constant of inertia for each machine and the oscillation modes. The sum of the oscillation kinetic energy over all frequencies and all machines, which is the system oscillation energy, can be used as a system wide dynamic performance index because of its intuitive physical interpretation and convenient representation in terms of all modes \cite{Silva2017}. Consider the linearized power system equations with system matrix $A$, $p$ synchronous generators and $n$ state variables. By similarity transformation, the state variable vector $\Delta x$ and the  transformed state variable vector $\Delta z$ are related as $\Delta x=M \Delta z$, where $M=\{v_1,v_2,...v_n\}$ is the matrix of right eigenvectors, $\Lambda=M^{-1}AM=\diag \{\lambda_i\}$, and $\lambda_i$ is the i-th system eigenvalue. Thus, for a given initial value $\Delta x(0)=\Delta x_0$ at time $t=0$:
\begin{align}
\left.
\begin{array}{l}
\Delta \dot{x} = A\Delta x\\
\Delta x(0) = \Delta x_0\\
\end{array}
\right\}
\Rightarrow
\begin{array}{l}
\Delta \dot{z} = \Lambda \Delta z =\Lambda \Delta z\\
\Delta z(0) = \Delta z_0=M^{-1} \Delta x_0\\
\end{array}
\label{eq:sol}
\end{align}
The system oscillation energy is as follows:
\begin{align}
E_k(t)=&\sum_{j=1}^{p}  \frac{1}{2}J_j\Delta \omega^2_j  =\frac{1}{2}\Delta x^T J \Delta x\\
  =&\frac{1}{2}\Delta z^T G\Delta z  \in \mathbb{R}
\label{eq:ke}
\end{align}
where $J$ and $G=M^TJM$ are the inertia and the transformed inertia matrices, respectively (see reference \cite{Silva2017}), and $\omega_j$ is the rotational speed in per unit of jth generator. Consider now the system action ($S$), which is the integral over time of the system kinetic energy:
\begin{align}
S(\tau)=& \int_{0}^{\tau} E_k(t)dt=\int_{0}^{\tau} \frac{1}{2}(\Delta z^T G\Delta z)dt \in \mathbb{R}\\
S(\tau) =&\frac{1}{2}\sum_{j=1}^n\sum_{i=1}^n\frac {e^{(\lambda_i+\lambda_j)t}}{(\lambda_i+\lambda_j)} z_{0i}z_{0j}g_{ij}\bigg\rvert_{0}^{\tau}
\label{eq:D}
\end{align}
where $z_{0i}$ is the i-th element of $\Delta z_0=M^{-1}\Delta x_0$ and $g_{ij}$ is the entry in the i-th row and j-th column of $G$. Assuming stability, the total action is defined as,
\begin{align}
S_{\infty} =&\lim_{\tau \to \infty} S(\tau) = -\frac{1}{2}\sum_{j=1}^n\sum_{i=1}^n\frac {z_{0i}z_{0j}g_{ij}}{(\lambda_i+\lambda_j)}
\label{eq:DsolT}
\end{align}
Note that $E_k \rightarrow 0$ as $t \rightarrow \infty$ and $E_k(t)>0,~\forall~t$. The best dynamic performance occurs when $E_k$ quickly approaches zero, which is equivalent to the case when the total action is minimized.

\subsection{Linear estimation of the total action}
Consider eigenvalue displacements caused by changes in an operational parameter such as the injected wind power $\Delta P_w  =P_w-P_{w0}$, a random variable. As the total action must be calculated under the new operating conditions after the changes, to reduce computational burden, an approximation of the total action as a function of the random variable is of interest. By linearizing Equation \eqref{eq:DsolT} around initial eigenvalues $\lambda_i^0$, the following expression is obtained:
 \begin{align}
\Delta S_\infty &\approx \sum_{i=1}^n \frac{\partial S_\infty}{\partial \lambda_i} \frac{\partial \lambda_i}{\partial P_w} \Delta P_w= \underbrace{\left ( \sum_{i=1}^n \beta_i \frac{\partial \lambda_i}{\partial P_w}\right )}_\gamma \Delta P_w
\label{eq:delta_s}
\end{align}
where
\begin{align}
&\beta_i = \frac{\partial S_\infty}{\partial \lambda_i} = \sum_{j=1}^n \frac{z_{0i}z_{0j}g_{ij}}{(\lambda_i^0+\lambda_j^0)^2}\\
&\frac{\partial \lambda_i}{\partial P_w} = l_i^T \frac{\partial A}{\partial P_w} v_i
\end{align}
Here $l_i$ and $v_i$ are the left and right column-eigenvectors associated with $\lambda_i$, respectively. Note that $\Delta S_{\infty}$ is a real number, although $\beta_i$ and $\partial \lambda_i / \partial P_w $ are all complex quantities. Based on preliminary evaluations, for an important range of operating conditions, the terms $\partial \lambda_i / \partial P_w $ can be assumed constant and equal to those calculated at the initial condition.

Assume now that we are interested in comparing the total action when different EIR locations are chosen to improve the system oscillations under wind power variability. If $k$ is the bus where the EIR is placed, the total action becomes:
 \begin{align}
S_\infty^k &\approx  S_\infty^{0k} +\gamma_k \Delta P_w
\label{eq:linear}
\end{align}
where $S_\infty^{k}$ is the total action estimated for EIR at bus $k$, $S_\infty^{0k}$ is the initial total action for bus k at the initial wind power $P_{w0}$ and $\gamma_k$ is the linear coefficient of equation \eqref{eq:delta_s} when EIR is connected at bus $k$. 

\section{Chance constrained optimization} \label{sec:stochastic}

\subsection{Formulation}
Because the wind power injection $\Delta P_w$ is a random variable, the system matrix $A$, $\Delta \lambda_i ~ \forall ~ i$ and the total action in Equation \eqref{eq:DsolT} become random variables as well. Conditional probabilities associated with the total action can be computed for a given disturbance. By modeling $x_0=\Delta x_0$ as a random variable, and obtaining some probability density function of common disturbances based on system data, the total probability of the total action can be calculated using Bayes' rule. Using this information and recalling that the total action works as a system dynamic performance measure regarding oscillations, what is the system location where an EIR based damping control should be installed? This DRP problem is stated as:

\begin{align}
k^*= \argmax_{k \in \mathcal{K}}~ \Phi_k\label{eq:opt}
\end{align}
where,
\begin{align}
\Phi_k&= \sum_{x_0 \in X_0} \prob (S_\infty^k = \min_{i \in \mathcal{K}} S_\infty^i|x_0) \times \prob(x_0)\\
&=\sum_{x_0 \in X_0}\prob (S_\infty^k \leq S_\infty^1 ...\cap S_\infty^k \leq S_\infty^m |x_0)\times \prob(x_0)
\end{align}
%
Here $X_0$ is the set of initial disturbances and $\mathcal{K}$, with $|\mathcal{K}|=m$, is the set of bus candidates to connect the EIR. Algorithm \ref{Alg:TR} shows the procedure to solve this chance-constrained optimization using the linear estimation of the total action.

\begin{algorithm}[!h]
\caption{Chance constrained optimization} \label{Alg:TR}
\begin{algorithmic}[1]
\State Get random vector with $N$ number of samples $P_w$
\For{each disturbance $x_0 \in X_0$}
\For{each actuator $k \in \mathcal{K}$}
\State Calculate $\gamma_k$ and $S_\infty^{0k}$
\State Compute $S_\infty^k$ according to \eqref{eq:linear}
\EndFor
\For{each actuator $k \in \mathcal{K}$}
\State Determine $\prob_k=\prob (S_\infty^k = \min_{i \in \mathcal{K}} S_\infty^i|x_0)$
\State Update $\Phi_k^{new}$ = $\Phi_k^{old} + \prob_k\times \prob(x_0)$
\EndFor
\EndFor
\State Obtain $k^*$ according to \eqref{eq:opt}
\end{algorithmic}
\end{algorithm}

\section{Case study}

\begin{figure}[ht!]
\centering
\includegraphics[scale=2.6]{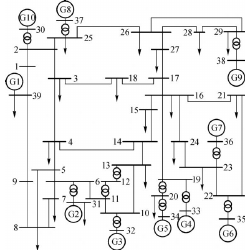}
\caption{IEEE 39-bus test system}
\label{fig:new_england}
\end{figure}

The IEEE 39-bus system is employed for simulations. Each synchronous generator is represented by a 6th order model with an IEEE type-1 exciter and an IEEEG1 governor. The wind power variability is studied by adding an equivalent 1,000 MW wind turbine connected at bus 16 in Figure \ref{fig:new_england} ($\approx$ 30\% of the system load). The equivalent wind turbine is modeled as a static generator with fixed dispatched power, i.e., no additional dynamics of the wind turbine are included. The set of generator buses $\mathcal{K} = \{39,~31,~32,...,~30\}$ corresponds to the set of bus candidates to connect the EIR. For simulation the EIR used in this paper corresponds to a 200 MW ESS (battery). The damping control consists of a proportional gain between the frequency changes at the connection bus and the reference active power. Data for controllers, system parameters and the full battery model are obtained from the simulation library in DIgSILENT PowerFactory.

\subsection{Preliminary deterministic analysis}
\begin{figure}[ht!]
\centering
\includegraphics[width=0.85\columnwidth]{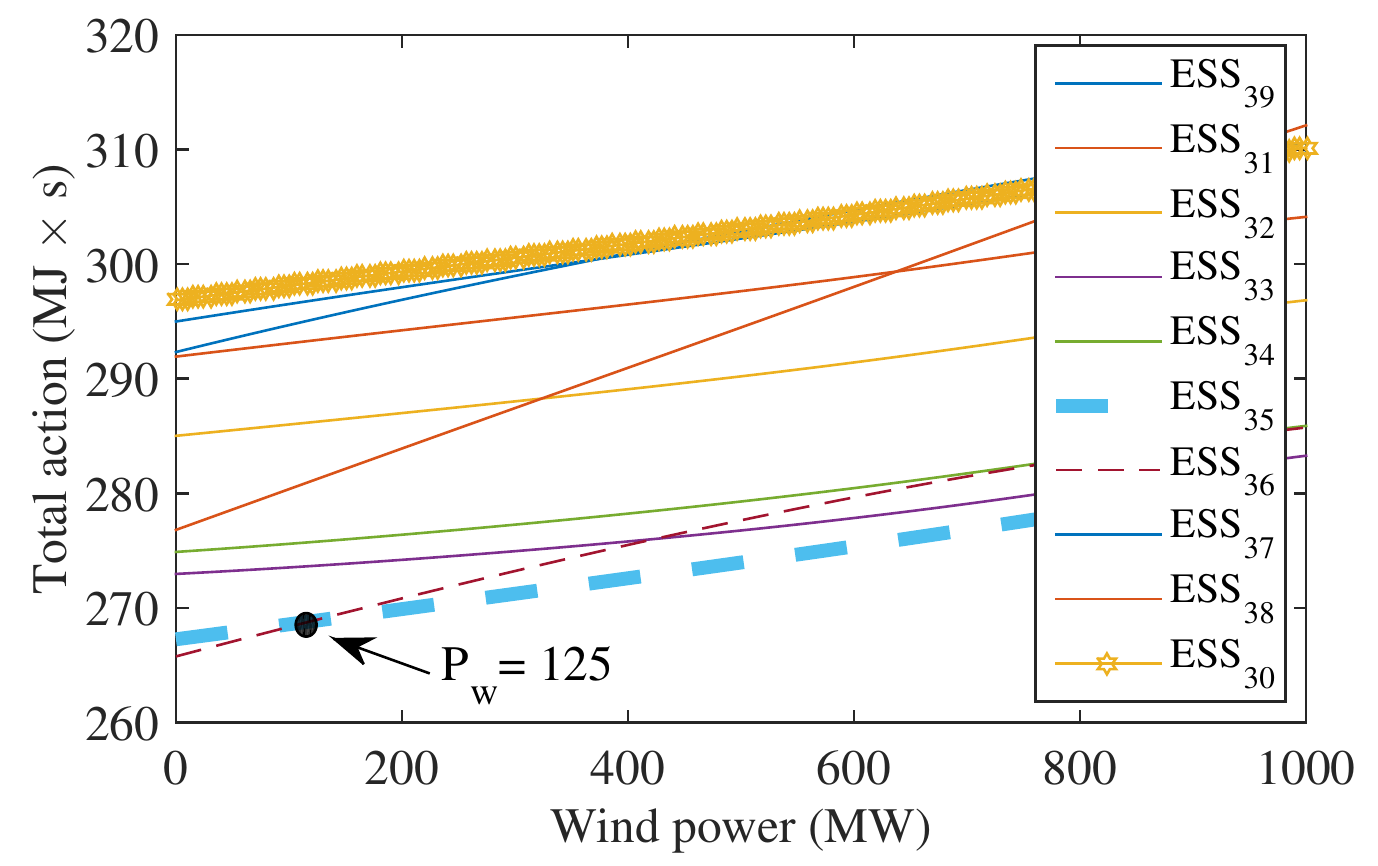}
\caption{Parameterized total action for different ESS location.}
\label{fig:TA}
\end{figure}

To first get insights of the chance-constrained optimization stated in section \ref{sec:stochastic}, a parameterized analysis is performed to calculate the exact total action for a disturbance in machine speed $\omega_1$---a short circuit at bus 39. The injected power of the wind turbine is used as a parameter, which is varied from 0 to 1,000 MW.
For each prospective ESS location, the parameterized analysis is performed. At each value of wind power generation, the power system equations are linearized, eigenvalues and right/left eigenvector of the system matrix $A$ calculated, and the total action defined in Equation \eqref{eq:DsolT} determined. Figure \ref{fig:TA} shows the total action as a function of the wind power for each ESS$_k$ connected at bus $k$.

\begin{figure}[ht!]
\centering
\includegraphics[width=0.85\columnwidth]{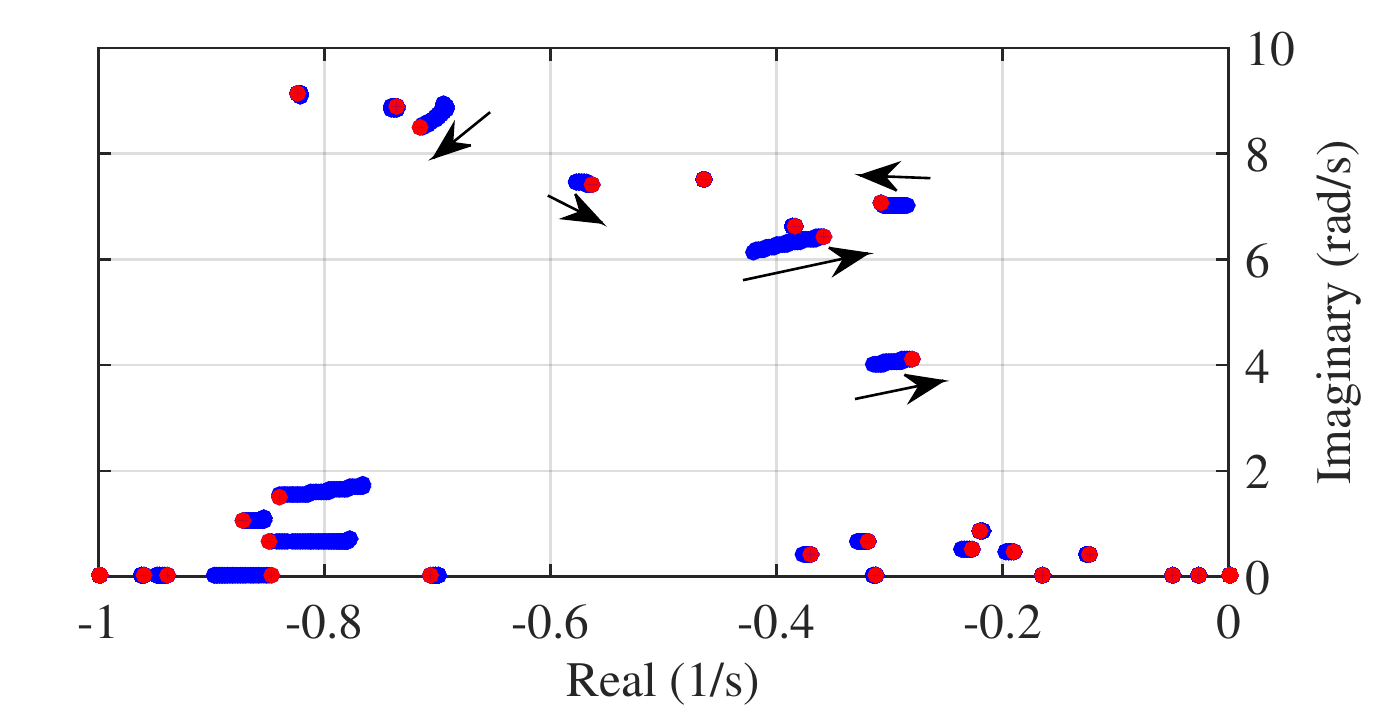}
\caption{Eigenvalue trajectories for the parameterized case without ESS.}
\label{fig:evals}
\end{figure}

The results show that for $P_w = 0$ the best bus to place an ESS-based damping control corresponds to bus 36---bus of generator 7. This means that when the ESS is connected to bus 36 the total action is minimized, providing the optimal dynamic performance for system oscillations. However, as the wind power increases, the solution changes and bus 35 becomes the best location---bus of generator 6. Therefore, the optimal location for installing the ESS depends on the wind power, which is a random variable. For this particular disturbance, $\prob(P_w<125~MW)$ gives the probability that the optimal solution is given by connecting the ESS at bus 36. Similarly, different disturbances can provide different solutions. This shows that the chance-constrained optimization in \eqref{eq:opt} is needed to solve the DRP problem. Figure \ref{fig:TA} is obtained using the exact calculation of the total action for the ESS connected at each bus at a time. This calculation requires significant computational resources, and therefore, it is impractical in real systems. Consider instead the linear approximation in Equation \eqref{eq:linear}.  Note that this estimate assumes eigenvalue trajectories to be linear with respect to changes in the wind power. Figure \ref{fig:evals} verifies this assumption, which, due to space limitations, focuses only the base case without ESS. The arrows in Figure \ref{fig:evals} represent the direction of eigenvalue trajectories for the electromechanical modes when the parameter $P_w$ is increased. 

\subsection{Stochastic analysis}

\begin{figure}[ht!]
\centering
\includegraphics[width=0.85\columnwidth]{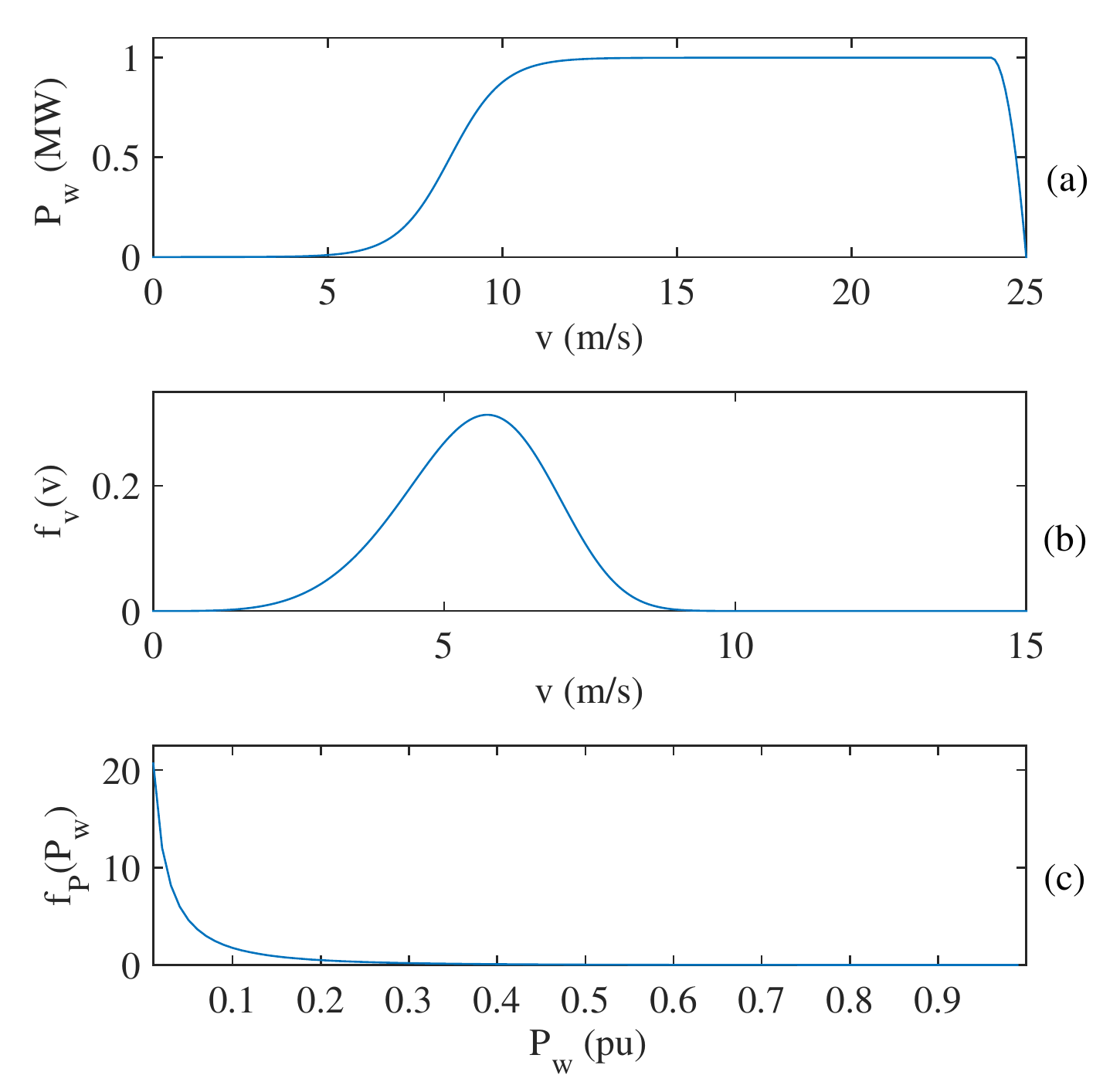}
\caption{(a) Wind speed-power characteristic, (b) Weibull distribution of the wind speed, (c) pdf of the wind power.}
\label{fig:wind}
\end{figure}

The chance-constrained DRP problem is solved through Monte Carlo simulations using the linear approximation of the total action for an initial operation at $P_{w0}=0$. The sample points $P_w$ are obtained from the pdf of the wind power generated by a Weibull distribution of wind speed. Figure \ref{fig:wind} shows (a) the deterministic relationship between wind speed and power,  (b) the probability density function of the wind speed, and (c) the probability density function of the wind power. The latter is obtained by random variable transformation; parameters can be found in \cite{Dhople2012}.

\begin{figure}[ht!]
\centering
\includegraphics[width=0.85\columnwidth]{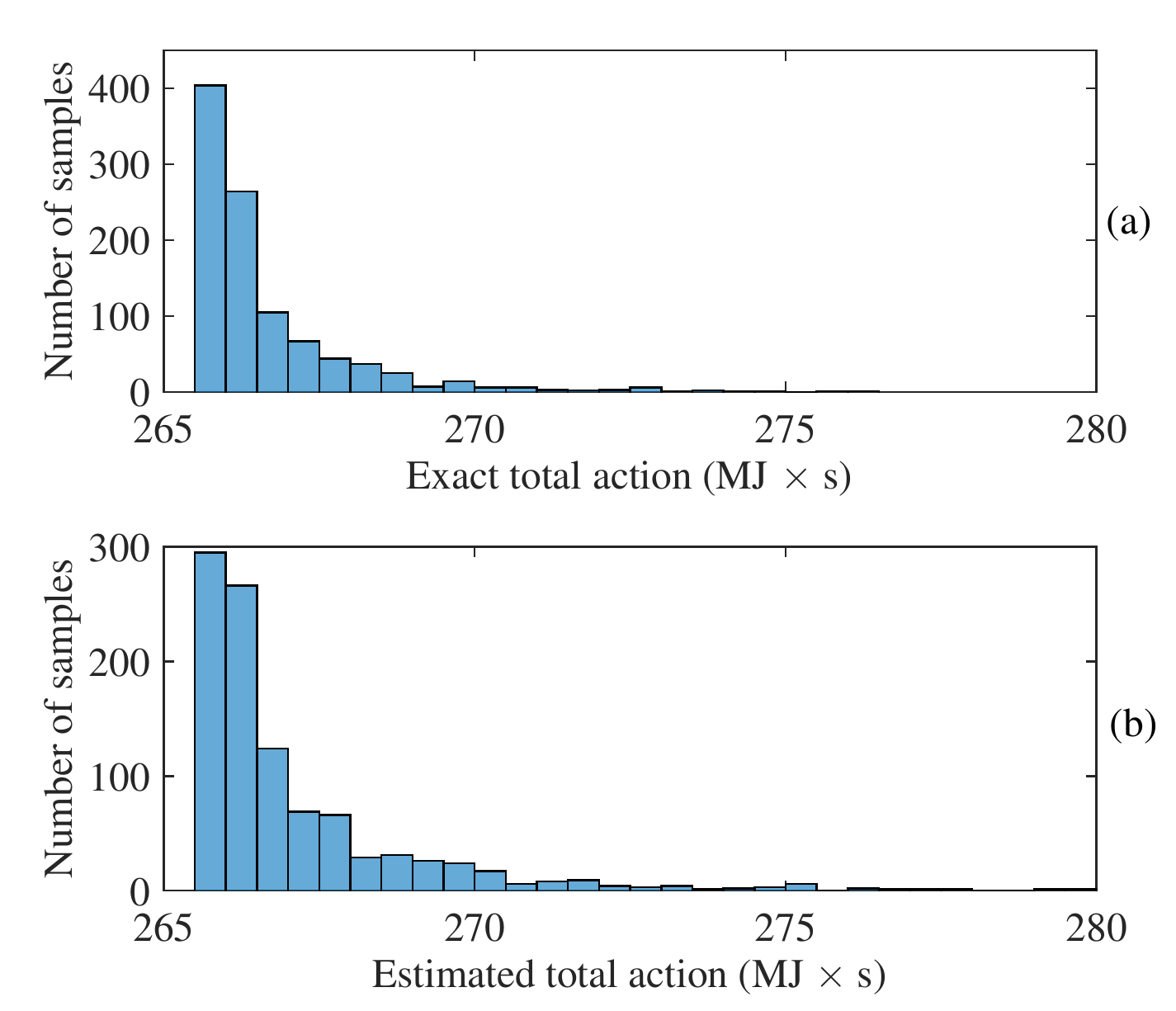}
\caption{Histograms of total action when the ESS is connected at bus 36: (a) exact total action, (b) estimated total action.}
\label{fig:histogram}
\end{figure}

Before applying Algorithm \ref{Alg:TR}, a total of $1,000$ sample points were employed to compare the results from the exact and estimated calculation of the total action. Figure \ref{fig:histogram} shows the histograms of the total action when the ESS is connected at bus 36 for a disturbance in the speed of generator 1.
This preliminary result shows agreement between the distributions of the exact and approximate total action. The big advantage of the approximation is the reduction in computational time, which is about 2 minutes for all the bus cansidates---including the calculation of $\gamma_k$---while the exact calculation computing eigenvalues is around 4 hours using an  Intel\textregistered Core\texttrademark i7-4790 CPU @3.6 GHz processor. Consider now $N = 10^6$ sample points of $P_\omega$ to solve the chance-constrained problem using the linear estimation, and the set  of equally likely disturbances $X_0 = \{x_0^1,~x_0^2,...,~x_0^7\}$, where $\Delta \omega_i = 0.01$ for each disturbance case. Table \ref{tab:disturbances} shows the disturbances and the probabilities $P_k$, which corresponds to the probability of the ESS at bus $k$ providing the minimum total action among all candidates buses over all wind power scenarios.


\begin{table}[h!]
\renewcommand{\arraystretch}{1.3}
\caption{Disturbances and probabilities of optimal location for each disturbance}
\label{tab:disturbances}
\centering
\begin{tabular}{c|c|c}
\hline
\hline
 & $\Delta \omega_i$ & $\prob (S_\infty^k = \min S_\infty^i)$\\
\hline
$x_0^1$ & $\Delta \omega_1$ &  $\prob_{35}=0.1$, $\prob_{36}=0.9$\\
$x_0^2$ & $\Delta \omega_1,\Delta \omega_8,\Delta \omega_{10}$ & $\prob_{35}=0.07$, $\prob_{36}=0.93$\\
$x_0^3$ & $\Delta \omega_1,\Delta \omega_2,\Delta \omega_{3},\Delta \omega_{4},\Delta \omega_{5}$ & $\prob_{32}=0.05$, $\prob_{36}=0.95$\\
$x_0^4$ & $\Delta \omega_8,\Delta \omega_9,\Delta \omega_{10}$ & $\prob_{38}=0.01$, $\prob_{30}=0.99$\\
$x_0^5$ & $\Delta \omega_1,\Delta \omega_2,\Delta \omega_{3},\Delta \omega_{8},\Delta \omega_{10}$ & $\prob_{35}=0.05$, $\prob_{36}=0.95$\\
$x_0^6$ & $\Delta \omega_2,\Delta \omega_3$ & $\prob_{32}=1$\\
$x_0^7$ & $\Delta \omega_9$ &$\prob_{38}=1$\\
\hline
\hline
\end{tabular}
\end{table}

Using the results given in Table \ref{tab:disturbances}, the objective function $\Phi_k$ in \eqref{eq:opt} can be computed. Table \ref{tab:results} shows the results.
\begin{table}[h!]
\renewcommand{\arraystretch}{1.3}
\caption{Results of the probability that each bus provides the best dynamic reinforcement.}
\label{tab:results}
\centering
\begin{tabular}{ccccc}
\hline
\hline
$\Phi_{39}$&$\Phi_{31}$&$\Phi_{32}$&$\Phi_{33}$&$\Phi_{34}$\\
$0$    &  $0$    & $0.15$ &   $0$  &$0$ \\
\hline
$\Phi_{35}$&$\Phi_{36}$&$\Phi_{37}$&$\Phi_{38}$&$\Phi_{30}$\\
$0.03$  &$0.53$  &$0$     &$0.14$&$0.14$\\
\hline
\hline
\end{tabular}
\end{table}
Note that the chance-constrained optimization determines that bus 36, terminal bus of generator 7, provides the optimal solution for the DRP problem. Thus, connecting an ESS at bus 36 enhances the system dynamics with the highest probability ($\Phi_{36}=0.53$ ). In order to show the superiority of the approach proposed in this paper, different comparison with traditional stochastic analysis are performed. Table \ref{tab:comparison} summarizes the results of the best ESS location obtained by: (a) maximizing the probability of having the dominant mode with a damping ratio above the benchmark, (b) maximizing the probability of having all damping ratios above the benchmark and (c) maximizing the probability of total action being the minimum among all bus candidates. The original system without ESS and with $P_w=0$ has critical damping ratios between 4 and 8\%, therefore, the benchmark is chosen to be 5\%. The results in Table \ref{tab:comparison} show that the traditional methods based on arbitrary benchmarks lead to an inefficient solution ($\Phi_{30}<\Phi_{36}$). In this case, the dominant mode corresponds to a local oscillation related to generator 10 connected at bus 30. An ESS connected at bus 30 will improve the local oscillation, however, the reduction in the energy of this oscillation does not improve overall system dynamics much and the probability of exciting this oscillation is small, which creates idle resources regarding the damping capacity of the ESS. The improvement of such local oscillation can be solved locally and does not require a system planning stage. Summarizing, solutions based on arbitrary displacement of eigenvalues may be operationally sufficient, but they are not optimal. On the other hand, the total action and disturbance based chance-constrained optimization guarantees the best dynamic behavior.
%
\begin{table}[h!]
\renewcommand{\arraystretch}{1.3}
\caption{Methods comparison}
\label{tab:comparison}
\centering
\begin{tabular}{c|c}
\hline
\hline
Optimization method & Optimal solution \\
\hline
(a) Benchmark for dominant mode & $k = 30$ \\
(b) Benchmark for all modes & $k = 30$ \\
(c) Total action and disturbance based& $k = 36$ \\
\hline
\hline
\end{tabular}
\end{table}

\section{Conclusion}

This paper describes a novel approach to guarantee the optimal EIR installation location considering dynamic reinforcement under wind power variability. The proposed chance-constrained optimization uses an energy-based index---total action---to measure system dynamic performance by combining all system eigenvalues rather than the study of a dominant mode or an arbitrary operational benchmark for damping ratios. Additionally, this method benefits from the treatment of disturbance probabilities. Simulations are implemented in the IEEE-39 bus system with 30\% of wind penetration, and the location of an ESS is analyzed with a linear estimation of the total action. Results show the the linear estimation drastically reduces the computation time from 4 hours to 2 minutes. Moreover, the comparison of the results with traditional approaches demonstrates its superiority by choosing a location that maximizes the probability of having the best performance, while the solutions obtained by other methods lead to inefficient installations.

\bibliographystyle{IEEEtran}
\bibliography{bare_conf}

\end{document}